\documentclass{article}

\usepackage{arxiv}

\usepackage[utf8]{inputenc} 
\usepackage[T1]{fontenc}    
\usepackage{hyperref}       
\usepackage{url}            
\usepackage{booktabs}       
\usepackage{amsfonts}       
\usepackage{nicefrac}       
\usepackage{microtype}      
\usepackage{graphicx}
\usepackage{doi}
\usepackage{algorithmic}
\usepackage{bm}
\usepackage{amsmath}
\usepackage{stmaryrd}
\usepackage{amssymb}

\title{Entropy-stable fluxes for high-order Discontinuous Galerkin simulations of high-
enthalpy flows}

\author{\href{https://orcid.org/0000-0002-1434-7166}{\includegraphics[scale=0.06]{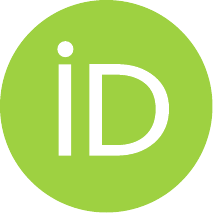}\hspace{1mm}Georgii Oblapenko$^1$}\thanks{
Preprint submitted to 
Notes on Numerical Fluid Mechanics and Multidisciplinary Design
},
{Arseniy Tarnovskiy$^1$},
\href{https://orcid.org/0000-0002-1900-5122}{\includegraphics[scale=0.06]{orcid.pdf}\hspace{1mm}Moritz Ertl$^2$},
\href{https://orcid.org/0000-0002-1434-7166}{\includegraphics[scale=0.06]{orcid.pdf}\hspace{1mm}Manuel Torrilhon$^1$}\\
	$^1$Chair of Applied and Computational Mathematics, RWTH Aachen,
	Schinkelstrasse 52, 52062 Aachen, Germany \\
$^2$Institute of Aerodynamics and Flow Technology, German Aerospace Center (DLR),\\
37073 G\"{o}ttingen, Germany\\
	$^\ast$Corresponding author. E-mail: \texttt{oblapenko@acom.rwth-aachen.de}
}

\renewcommand{\shorttitle}{Entropy-stable fluxes for high-order Discontinuous Galerkin simulations of high-
enthalpy flows}

\hypersetup{
pdftitle={vsc},
pdfsubject={},
pdfauthor={Georgii Oblapenko, Arseniy Tarnovskiy, Moritz Ertl, Manuel Torrilhon},
pdfkeywords={},
}

\begin{document}
\maketitle

\begin{abstract}In the present work, we extend the Discontinuous Galerkin Spectral Element Method (DGSEM) to high-enthalpy
reacting gas flows with internal degrees of freedom. An entropy- and kinetic energy-preserving
flux function is proposed which allows for use of arbitrary expressions for the internal energies
of the constituent gas species.
The developed method is applied to simulation of several model problems and compared to the DLR TAU solver.
\end{abstract}


\section{Introduction}
Accurate simulations of high-temperature reacting multi-species flows are a crucial component of the research and development cycle
in multiple industrial and scientific applications, such as combustion and spacecraft design~\cite{karl2024sustainable}.
As available computing power increases, so does the range of physical phenomena that can be modelled. At the same time, development of new numerical methods leveraging modern computing architectures is a crucial part of CFD development~\cite{wang_high-order_2013,slotnick2014cfd}. Discontinuous Galerkin (DG) methods possess many properties attractive for next-generation CDF solvers~\cite{shu2013brief,wang_high-order_2013}: 1) higher-order solution representation 2) low numerical dissipation 3) less stringent requirements on grid-shock alignment 4) improved computational efficiency. At the same time, their higher-order nature and insufficient numerical dissipation may lead to numerical stability issues~\cite{gassner2021novel}.

Therefore, recent focus has been on the development of DG schemes with provable stability and structure-preserving properties, such as entropy conservation and/or entropy-stability~\cite{tadmor2003entropy,chan2018discretely,chen2020review,gassner2021novel}, and provably entropy-stable and positivity-preserving limiting approaches~\cite{hennemann2021provably,rueda2022subcell}.

However, extending these methods to multi-species high-temperature reacting flows with real gas effects is challenging due to the intricate expressions for the internal energy and specific heats, making entropy-conservative fluxes harder to derive and compute. Examples of application can be found
in~\cite{gouasmi2020formulation,johnson2020conservative,ching2024positivity}, where internal energies were  modelled using NASA polynomials~\cite{mcbride2002nasa}, and in~\cite{osti_1763209,peyvan2023high}, where the simple infinite harmonic oscillator model was used to model the vibrational spectra of the molecular species.

In our recent work~\cite{oblapenko2024entropy}, we proposed a new approach to the derivation of consistent entropy-conservative flux functions which does not enforce any restrictions on the internal energy functions, and whose computational cost is independent of 
the complexity of the expressions for the internal energies. The approach is based on linear interpolation of the specific heats
and internal energies using pre-computed tabulated values~\cite{hempert2017simulation} and exact integration for the entropy.
It has been shown to be kinetic energy-preserving and entropy-conserving and has been successfully applied to high-enthalpyy single-species flows.
In the present work we extend the approach to the multi-species case and apply it to the simulation of a weak blast wave, the spatially homogeneous chemical relaxation of a gas mixture, and a reacting two-species Mach 10
flow around a cylinder. We demonstrate excellent agreement with reference solutions and showcase the application of the developed approach for simulating reacting flows with strong shocks. 

\section{Governing equations}
We consider a two-dimensional flow of an inviscid mixture of $N_c$ gases.

The compressible Euler equations governing such a flow are given by
\begin{equation}
    \frac{\partial}{\partial t}\mathbf{u} + \frac{\partial}{\partial x}\mathbf{f}_x + \frac{\partial}{\partial y}\mathbf{f}_y = R^{\mathrm{chem}}.\label{eqns:euler-generic}
\end{equation}
Here $t$ denotes the time, $x$ and $y$ are the spatial coordinates, $\mathbf{u}$ is the vector of conservative variables, and $\mathbf{f}_x$, $\mathbf{f}_y$ are the inviscid fluxes.

The vector of conservative variables  $\mathbf{u} \in \mathbb{R}^4$ is given by
\begin{equation}
     \mathbf{u} = \left(\rho_1, \ldots, \rho_{N_c}, \rho v_x, \rho v_y, E \right)^{\mathrm{T}},
\end{equation}
where $\rho_c$ is the density of species $c$, $v_x$ and $v_y$ are the flow velocities in the $x$ and $y$ directions, and $E=\rho e=\rho \varepsilon^{\mathrm{int}} + \rho v^2 / 2$ is the total flow energy. Here, $e$ is the specific energy, and $\varepsilon^{\mathrm{int}}$ is the specific internal energy computed as
\begin{equation}
    \varepsilon^{\mathrm{int}} = \sum_c Y_c \varepsilon^{\mathrm{int}}_c,
\end{equation}
where $Y_c=\rho_c / \rho$ and $\varepsilon^{\mathrm{int}}_c$ are given by
\begin{equation}
    \varepsilon^{\mathrm{int}}_c = \frac{3}{2}kT + \varepsilon^{\mathrm{f}}_c,\quad
    \varepsilon^{\mathrm{int}}_c = \frac{3}{2}kT +\varepsilon^{\mathrm{rot}}_c(T) + \varepsilon^{\mathrm{vibr}}_c(T) + \varepsilon^{\mathrm{f}}_c
\end{equation}
for atoms and molecules, respectively.

Here $\varepsilon^{\mathrm{f}}_c$ is the energy of formation of species $c$, $\varepsilon^{\mathrm{rot}}_c(T)$ and $\varepsilon^{\mathrm{vibr}}_c(T)$ are the rotational and vibrational energies of species $c$, correspondingly.
In the present work, we assume that $\varepsilon^{\mathrm{rot}}_c(T)=kT$. 
The species' constant-volume specific heats are defined as $
    c_{v,c}(T) = \partial \varepsilon^{\mathrm{int}}_c / \partial T$.

The inviscid fluxes are given by
\begin{equation}
     \mathbf{f}_x = \left(\rho_1 v_x, \ldots, \rho_{N_c} v_x, \rho v_x^2 + p, \rho v_x v_y, (E+p)v_x \right)^{\mathrm{T}},
\end{equation}
\begin{equation}
     \mathbf{f}_y = \left(\rho_1 v_y, \ldots, \rho_{N_c} v_y, \rho v_x v_y, \rho v_y^2 + p, (E+p)v_y \right)^{\mathrm{T}}.
\end{equation}
Here $p$ is the pressure. We use the ideal gas law to relate pressure, density, and temperature: $p=nkT$, where $n$ is the number density, $k$ is the Boltzmann constant, and $T$ is the flow temperature.

The vector of chemical production terms is given as
\begin{equation}
    R^{\mathrm{chem}} = \left(R^{\mathrm{chem}}_{1},\ldots,R^{\mathrm{chem}}_{N_c},0,0,0\right)^{\mathrm{T}}.
\end{equation}

Finally, we define the physical entropy $s$ as
\begin{equation}
    s = \sum_c Y_c \int_0^T \frac{c_{v,c}(\tau)}{\tau}\mathrm{d}\tau - \sum_c Y_c \frac{k}{m_c} \ln \rho_c = \sum_c Y_c \eta_c(T) -  \sum_c Y_c \frac{k}{m_c} \ln \rho_c. \label{eq:entropy-def}
\end{equation}
Here $m_c$ is the molecular mass of chemical species $c$. We will hereafter refer to the $\int_0^T \frac{c_{v,c}(\tau)}{\tau}\mathrm{d}\tau $ as the ``integral part'' of the entropy of species $c$ and denote it by $\eta(T)_c$ to simplify the notation. We also define the mathematical entropy as the volume density $\mathfrak{s}=-\rho s$.
Based on this definition of entropy, we now derive an entropy-conservative flux for Eqns.~(\ref{eqns:euler-generic}).

\section{Entropy-conservative flux}
Following the procedure of~\cite{oblapenko2024entropy}, we obtain the following expressions for the entropy-conservative density, momentum, and energy fluxes in the $x$ direction (fluxes in the $y$ direction are obtained by analogy and are not presented here):
\begin{eqnarray}
  F^{\mathrm{num},x}_{\rho_c} = & {\left\{\!\left\{ \rho_c \right\}\!\right\}_{\mathrm{log}}}\left\{\!\left\{ v_x \right\}\!\right\},\:c=1,\ldots,N_c,\label{eq:flux_fx_rho} \\
  F^{\mathrm{num},x}_{\rho v_x}  = & \left\{\!\left\{ v_x \right\}\!\right\} \sum_c F^{\mathrm{num},x}_{\rho_c}   + \sum_c\frac{k}{m_c}\frac{ \left\{\!\left\{ \rho_c \right\}\!\right\}}{ \left\{\!\left\{ 1/T \right\}\!\right\}}, \\
  F^{\mathrm{num},x}_{\rho v_y}  =& \left\{\!\left\{ v_y \right\}\!\right\} \sum_c F^{\mathrm{num},x}_{\rho_c} , \\
  F^{\mathrm{num},x}_{E} = & \sum_c F^{\mathrm{num},x}_{\rho,c} \left(
  \left\{\!\left\{ T \right\}\!\right\}_{\mathrm{geo}}^2 \left(\frac{c_{v,c}\left(T^\ast\right)}{T^\ast} -\left\{\!\left\{ \frac{1}{T} \right\}\!\right\} c_{v,c}\left(T^{\ast\ast}\right) \right) \nonumber \right.  \\
  & + \left. \left\{\!\left\{ \varepsilon^{\mathrm{int}}_c \right\}\!\right\} - \frac{\left\{\!\left\{ v_x^2 \right\}\!\right\} +\left\{\!\left\{ v_y^2 \right\}\!\right\}} {2}\right) + \left\{\!\left\{ v_x \right\}\!\right\} F^{\mathrm{num},x}_{\rho v_x} +  \left\{\!\left\{ v_y \right\}\!\right\}  F^{\mathrm{num},x}_{\rho v_y} .\label{eq:flux-E_x}
\end{eqnarray}
Here the braces denotes various averaging operators between two states $a_-$ and $a_+$ between which the flux is computed:
\begin{equation}
  \left\{\!\left\{ a \right\}\!\right\} = \frac{1}{2}\left(a_- + a_+\right),\quad
  \left\{\!\left\{ a \right\}\!\right\}_{\mathrm{geo}} = \sqrt{a_- a_+},\quad
  \left\{\!\left\{ a \right\}\!\right\}_{\mathrm{log}} = \frac{a_+ - a_-}{\log a_+ - \log a_-}.
\end{equation}
The quantities $\frac{c_{v,c}\left(T^\ast\right)}{T^\ast}$, $c_{v,c}\left(T^{\ast\ast}\right)$ are defined as:
\begin{equation}
  \frac{c_{v,c}\left(T^\ast\right)}{T^\ast} = \frac{\left \llbracket \eta_c(T) \right \rrbracket}{\llbracket T \rrbracket},\quad
  c_{v,c}\left(T^{\ast\ast}\right) = \frac{\left \llbracket \varepsilon^{\mathrm{int}}_c \right \rrbracket}{\llbracket T \rrbracket }.\label{eq:jump-eint}
\end{equation}
Here $\llbracket a \rrbracket$ = $a_+ - a_-$ is the jump in the value of a flow variable $a$ between two points.
In general, no closed-form expression is available for $\eta_c(T)$, and only for specific cases of internal energy spectra models (NASA polynomials, infinite harmonic oscillator) is it possible derive analytical expressions for $\eta_c(T)$ required for the calculation of the entropy-conservative flux~(\ref{eq:flux-E_x}). However, if we pre-compute values of $c_{v,c}(T)$ and $\varepsilon^{\mathrm{int}}_c(T)$ over a sufficiently large range of temperatures with a small discretization step $\Delta T$ and use piece-wise linear reconstructions of these functions (a common practice in CFD solvers to reduce computational effort), it is possible to use this piece-wise linear representation of $\varepsilon^{\mathrm{int}}_c(T)$ in order to obtain a closed-form expression for $\eta_c(T)$:
\begin{equation}
  \eta_c(T) = \sum_{i=0}^{N-1} \eta_{c,i} + \left(c_{v,c,N} - \frac{(c_{v,c}(T) - c_{v,c,N}) T_{N}}{\Delta T}\right) \ln\left(\frac{T}{T_{N}}\right) + \left(c_{v,c}(T) - c_{v,c,N}\right).\label{eq:eta_compute}
\end{equation}
Here $c_{v,c,i}$ denotes a tabulated value of $c_{v,c}(T)$ computed at a tabulated temperature of $T_i$, $c_{v,c}(T)$ is a linearly interpolated value of $c_{v,c}$ at some temperature $T$. $N$ is defined as $\lfloor T - T_{\mathrm{min}} / \Delta T \rfloor$, where $T_{min}$ is the minimum temperature used for the tabulation. The values $\eta_{c,i}$ are the integrals of $c_{v,c}(T)/T$ computed over the $\Delta T$-sized intervals using a linear interpolation of $c_{v,c}(T)$ over the interval:
\begin{equation}
    \eta_{c,i} = \left(c_{v,c,i} - \frac{(c_{v,c,i+1} - c_{v,c,i}) T_{i}}{\Delta T}\right) \ln\left(\frac{T_{i+1}}{T_{i}}\right) + \left(c_{v,c,i+1} - c_{v,c,i}\right).
\end{equation}
The values of $\sum_{i=0}^{N} \eta_{c,i}$ can also be tabulated for $N=0,\ldots N_{\mathrm{max}}$, reducing the computation of $\eta_c(T)$ to obtaining a value from a look-up table and evaluating one logarithm. The presented entropy-conservative numerical flux is therefore computationally efficient, with the cost of its evaluation independent of the complexity of the expressions for the internal energies and specific heats, as those are required only for precomputation of the tabulated values.

\section{Numerical results}
The developed fluxes were implemented in Trixi.jl~\cite{schlottkelakemper2020trixi,ranocha2021adaptive}, an extendable framework for solving systems of hyperbolic equations using the DGSEM method. The strong stability preserving Runge-Kutta methods~\cite{gottlieb2005high} SSPRK43 was used for time integration. The Trixi.jl implementation of the presented fluxes, along with the input files for the simulations, has been made publicly available~\cite{oblapenko2024entropyconservativemsRepro}.

\subsection{Weak blast wave}
To verify the entropy conservation property of the developed flux, we consider the weak blast wave case adapated from~\cite{hennemann2021provably}. The initial conditions are given by
\begin{equation}
    \begin{bmatrix}
           \rho \\
           v_x \\
           v_y \\
           p
         \end{bmatrix} = \begin{bmatrix}
           0.341388 \\
           0.0 \\
           0.0 \\
           101325.0
         \end{bmatrix}\,\mathrm{if}\,\sqrt{x^2+y^2}>0.5;\quad 
    \begin{bmatrix}
           \rho \\
           v_x \\
           v_y \\
           p
         \end{bmatrix} = \begin{bmatrix}
           0.399117 \\
           102.5 \cos(\phi) \\
           102.5 \sin(\phi) \\
           126149.6
         \end{bmatrix}\mathrm{else}.
\end{equation}
The gas was taken to be an O$_2$/O mixture, with the molar fractions of both molecules and atoms equal to 50\%. The cut-off harmonic oscillator model~\cite{nagnibeda2009non,oblapenko2024entropy} was used to model the vibrational spectrum of the oxygen molecules, and no chemical reactions were considered. A domain of size $[-2, 2]\times[-2, 2]$~m was used, discretized by a uniform $64\times 64$ grid. No numerical dissipation or flux limiting was applied, so as to test the entropy conservation property of the flux.

\begin{figure}[h!]
  \centering
  \includegraphics[width=0.75\textwidth]{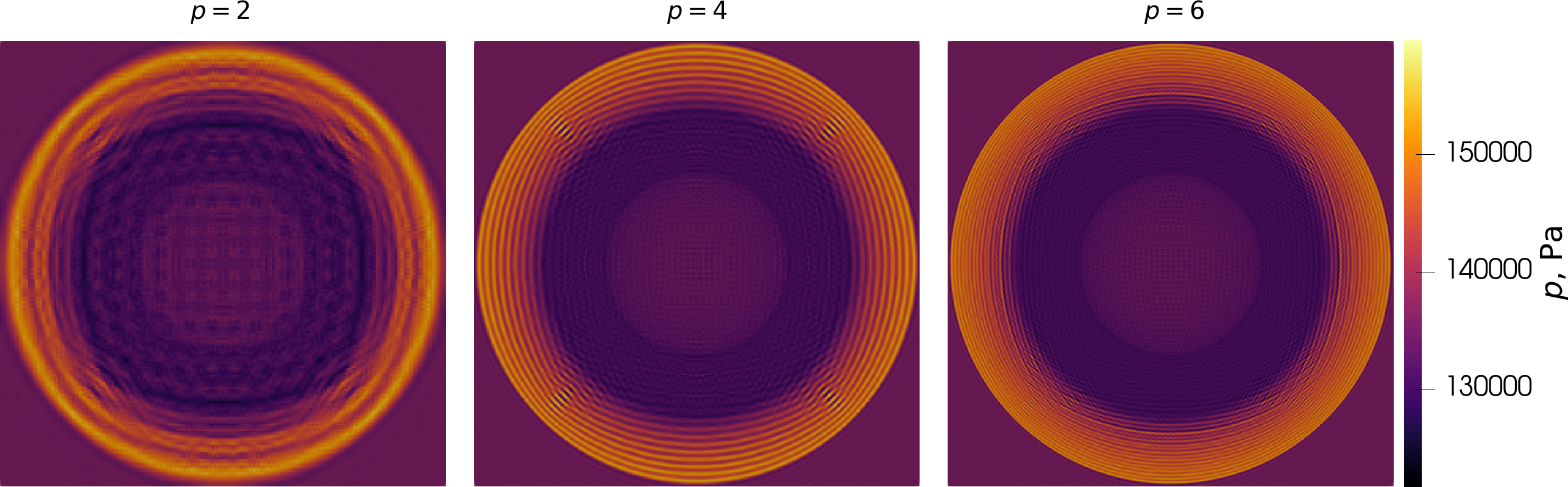}
  \caption{Pressure profile at $t=1.96$~ms for polynomials of degree 2, 4, and 6.}
  \label{fig:blast}
\end{figure}

Figure~\ref{fig:blast} shows the pressure at $t=1.96\cdot 10^{-3}$~s for three different orders of polynomials used: 2, 4, and 6. Due to the radial symmetry of the case and the coarse grid, significant artifacts can be seen for the polynomial degree of 2; use of higher-order polynomials alleviates this issue. Despite the numerical oscillations visible across the shock-wave front, appearing due to the absence of dissipation, the simulations run stable even without any limiting. 

\begin{figure}[h!]
  \centering
  \includegraphics[width=0.6\textwidth]{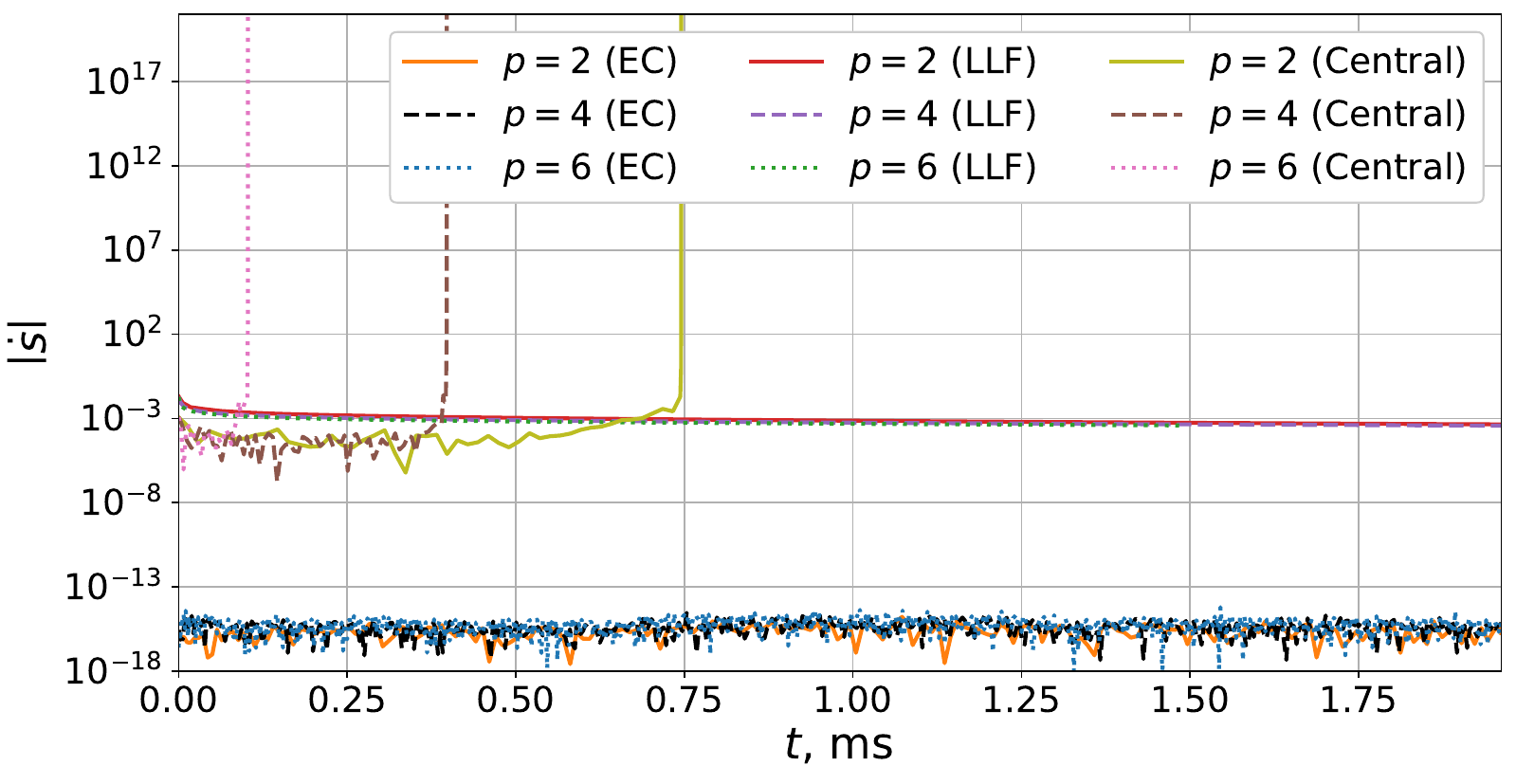}
  \caption{Entropy production rate as a function of time for different polynomial orders using the derived entropy-conservative flux (``EC''), the dissipative local Lax-Friedrichs flux (``LLF''), and the central flux (``Central'').}
  \label{fig:blast_srate}
\end{figure}

Figure~\ref{fig:blast_srate} shows the absolute entropy production rate plotted as a function of time over the course of the simulation. We see that for all polynomial orders considered, the entropy production is on the order of machine precision, thus confirming the entropy conservation property of the developed flux. As a comparison, the entropy production rates for simulations conducted using the dissipative local Lax-Friedrichs (LLF) flux and the non-dissipative central flux are also plotted. For the case of the  LLF flux, the entropy production rate can be seen to be 10--12 orders of magnitude higher than the negligibly small entropy production rate computed using the derived entropy-conservative fluxes. For the central flux, the simulations blow up, as an explicit time-stepping scheme is used.

\subsection{Spatially homogeneous chemical relaxation}
Next, we verify the implementation of chemical reactions. We consider the spatially homogeneous chemical relaxation of an O$_2$/O mixture, with the vibrational spectrum of O$_2$ modelled using the cut-off harmonic oscillator model. We consider only dissociation reactions, thus the chemical production terms are 
\begin{equation}
  R^{\mathrm{chem}}_{\mathrm{O}_2} = -m_{\mathrm{O}_2} n_{\mathrm{O}_2} n_{\mathrm{O}_2} k^{\mathrm{diss}}_{\mathrm{O}_2,\mathrm{O}_2} - m_{\mathrm{O}_2} n_{\mathrm{O}_2} n_{\mathrm{O}} k^{\mathrm{diss}}_{\mathrm{O}_2,\mathrm{O}}(T),\quad
  R^{\mathrm{chem}}_{\mathrm{O}} = -R^{\mathrm{chem}}_{\mathrm{O}_2}.
\end{equation}
The dissociation rates are computed using the Arrhenius law:
\begin{equation}
  k^{\mathrm{diss}}_{c,d}=A_{c,d}T^{n_{c,d}}\exp\left(-\frac{E_{D,c}}{kT}\right),
\end{equation}
where $E_{D,c}$ is the dissociation energy of the molecule of chemical species $c$, and $A_{c,d}$ and $n_{c,d}$
are the Arrhenius coefficients. In the present work, we use the following values of the Arrhenius coefficients~\cite{park1993review}:
$A_{\mathrm{O}_2,\mathrm{O}_2}=3.321 \cdot 10^{-9}$ m$^3$/s, $A_{\mathrm{O}_2,\mathrm{O}}=3.321 \cdot 10^{-9}$ m$^3$/s, 
$n_{\mathrm{O}_2,\mathrm{O}_2}=n_{\mathrm{O}_2,\mathrm{O}}=-1.6$.
Two different initial conditions were considered: 1) $T=8000$~K, $x_{\mathrm{O}_2}$=0.9, $n=10^{23}$ m$^{-3}$, 2) $T=12000$~K, $x_{\mathrm{O}_2}$=0.5, $n=10^{23}$ m$^{-3}$.

\begin{figure}[h]
  \centering
  \includegraphics[width=0.45\textwidth]{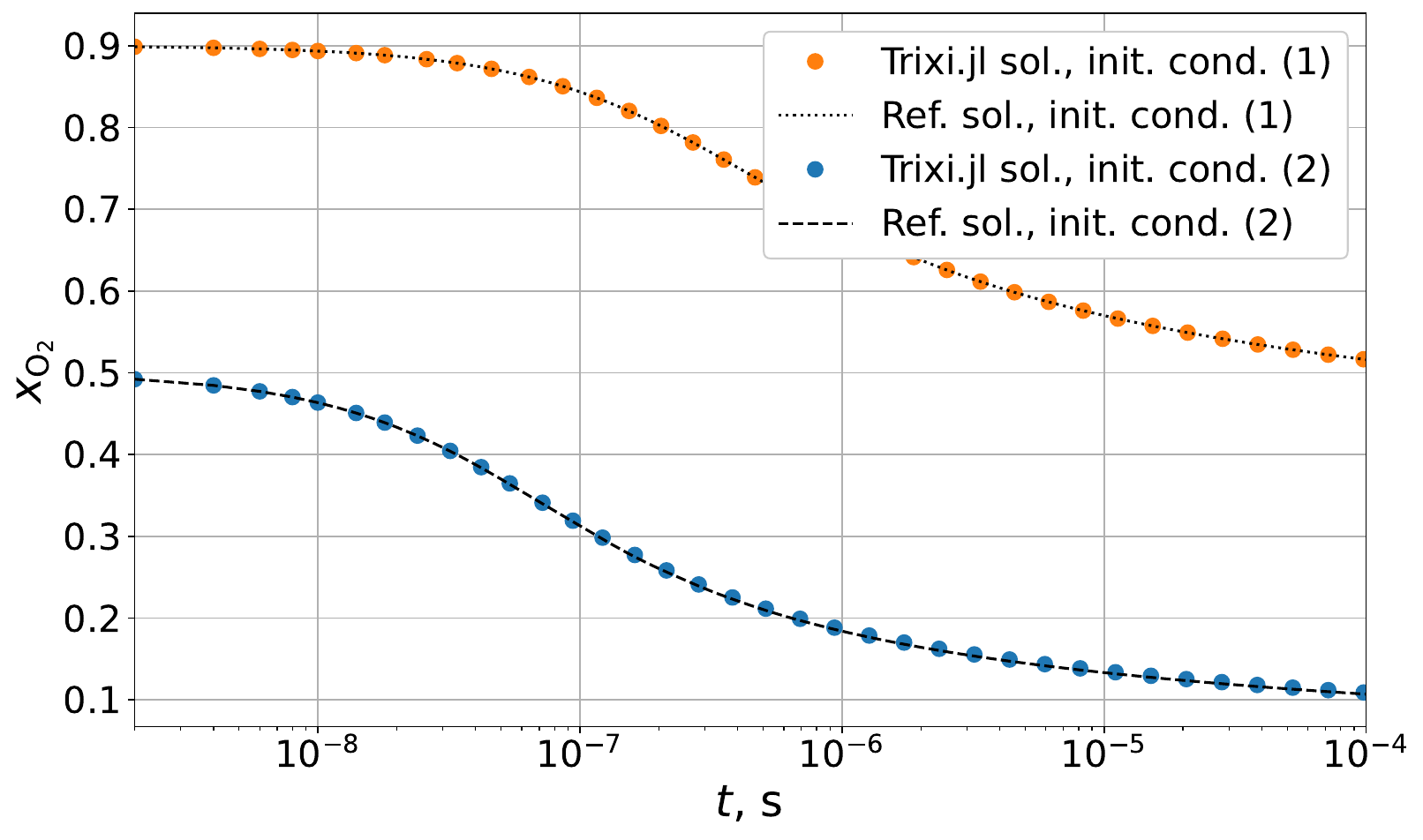}
  \includegraphics[width=0.4686\textwidth]{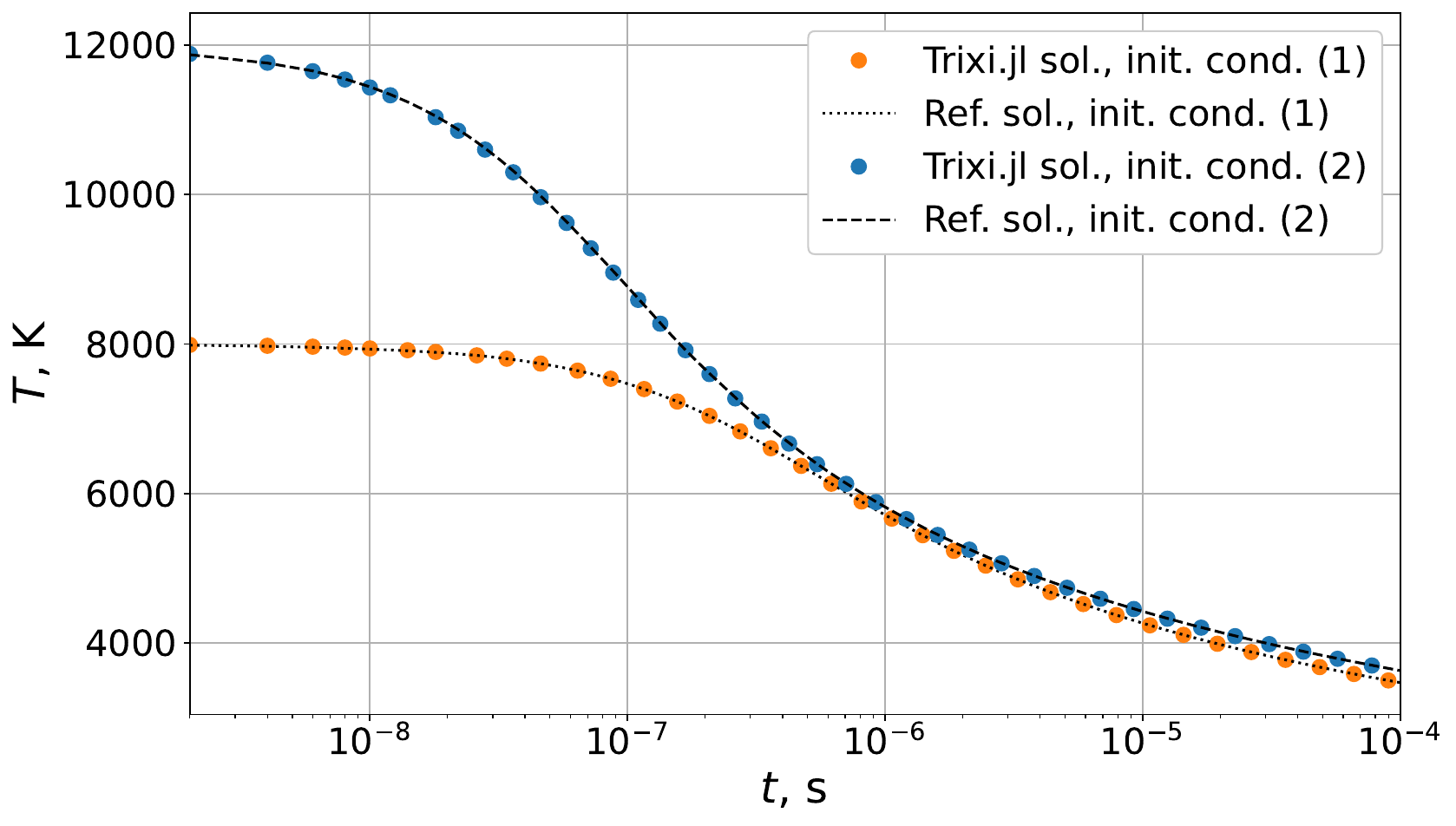}
  \caption{Temporal evolution of the molar fraction of O$_2$ (left) and temperature (right); lines show the reference solution
  of the system of master equations, whereas dots are the values computed via the Trixi.jl implementation.}
  \label{fig:xO2_evolution}
\end{figure}

Figure~\ref{fig:xO2_evolution} shows the evolution of the molar fraction of O$_2$ molecules and the gas temperature over time. The dots show to the solutions obtained via the Trixi.jl implementation, with orange dots corresponding to initial condition (1) and blue dots corresponding to initial condition (2). The dotted and dashed lines show the solution obtained by solving a system of master equations~\cite{nagnibeda2009non}, corresponding to initial conditions (1) and (2), respectively. We see excellent agreement between the implementation of chemical source terms in Trixi.jl and the reference solution, thus verifying our implementation of non-equilibrium chemistry.

\subsection{Mach 10 flow over a cylinder}
Finally, we consider a hypersonic flow of a reacting oxygen mixture around a cylinder. The same chemistry model was used as for the case of the spatially homogeneous chemical relaxation. For this test case, the cut-off anharmonic oscillator model was used to model the vibrational energy spectrum of the oxygen molecules. The free-stream conditions were taken as follows: $v_{x,\infty}=4000$~m/s, $p_{\infty}=450.2$~Pa, $T_{\infty}=400$~K, $x_{\mathrm{O}_2,\infty}=0.999733$.

The simulation was started on a 30$\times$30 grid with no refinement. The IDP subcell limiter~\cite{rueda2022subcell} was used; the SSPRK33 method was used for time integration.
Adaptive mesh refinement with up to 4 levels of refinement was used to improve the grid resolution near the shock using the capabilities of the p4est library~\cite{burstedde2011p4est}; the shock indicator from~\cite{hennemann2021provably} was used to mark cells for refinement. In total, approximately 3500 elements were used for the domain discretization. 3-rd and 4-th order polynomials were used for the simulations, leading to approximately 57000 and 86000 degrees of freedom per conservative variable field, respectively.

The DLR TAU solver~\cite{mack2002validation,hannemann2010closely} was used as a benchmark solver for this case with a similar initial grid and parameters governing the gas properties. The second order AUSM+ flux~\cite{liou1996sequel} with Green-Gauss gradient reconstruction was used, along with local time-stepping and a first-order implicit Backward Euler solver and a flux carbuncle fix \cite{Wada1994}. Six cycles of adaptive mesh refinement, with an increase in number of points of $30\%$ per cycle were performed in order to increase resolution at the shock until grid convergence. The resulting mesh consists of 8710 points.

\begin{figure}[h!]
  \centering
  \includegraphics[width=0.25\textwidth]{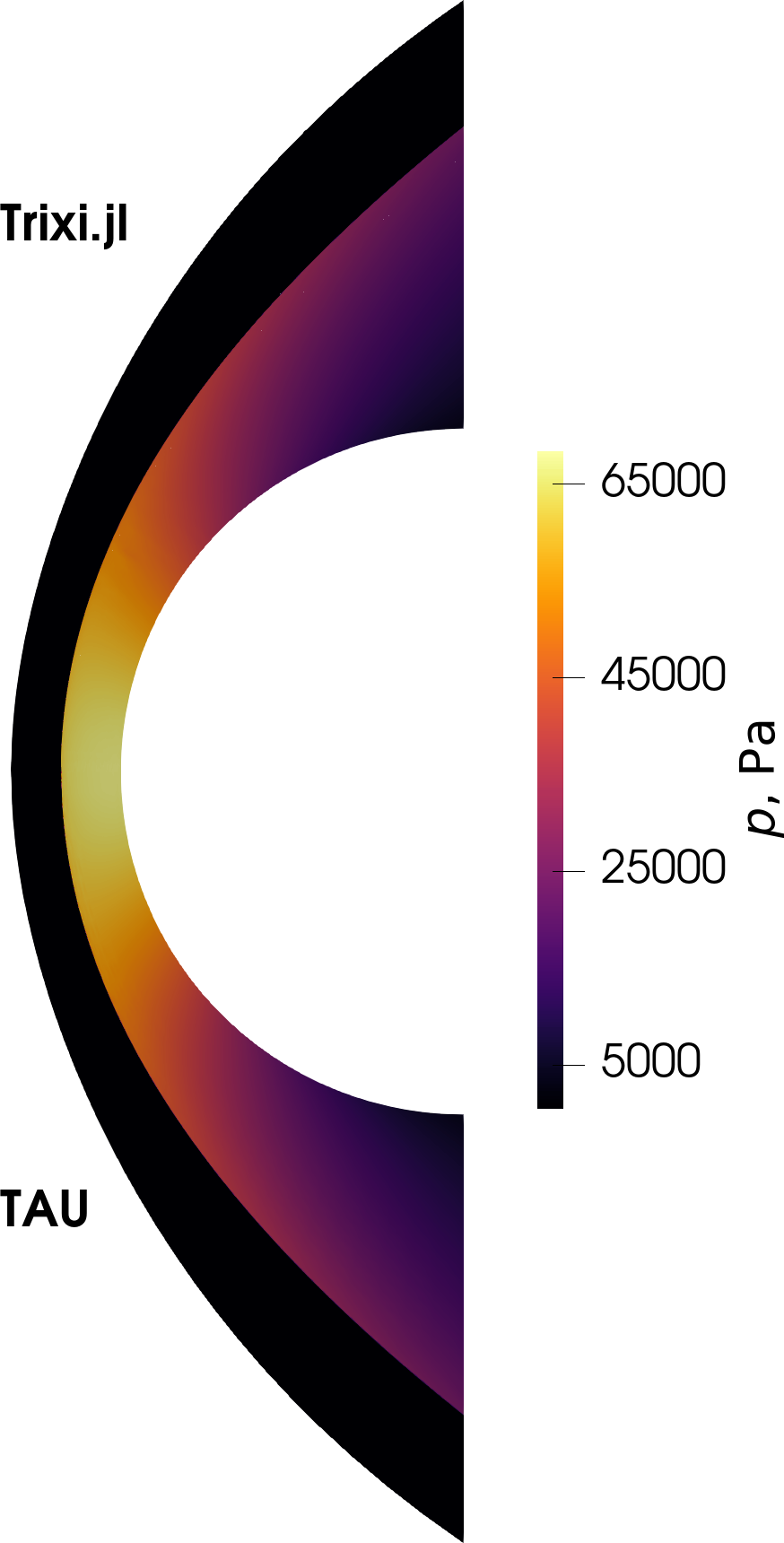}
  \includegraphics[width=0.2368\textwidth]{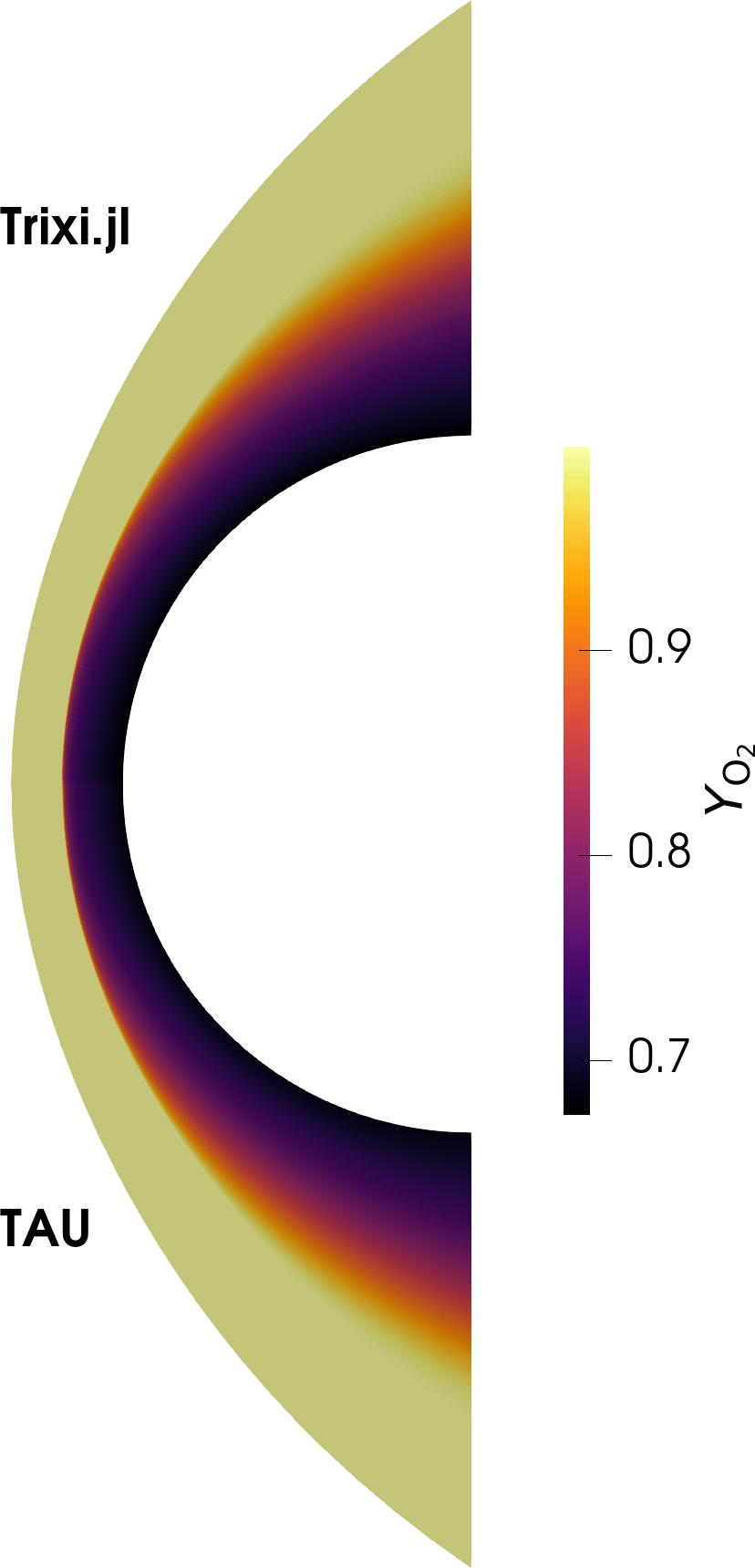}
  \caption{Pressure (left) and mass fraction of molecular oxygen (right) fields computed with a 3-rd order DG method with the entropy-conservative flux (top) and with the DLR TAU solver (bottom).}
  \label{fig:TrixiTau}
\end{figure}
\begin{figure}[h!]
  \centering
  \includegraphics[width=0.75\textwidth]{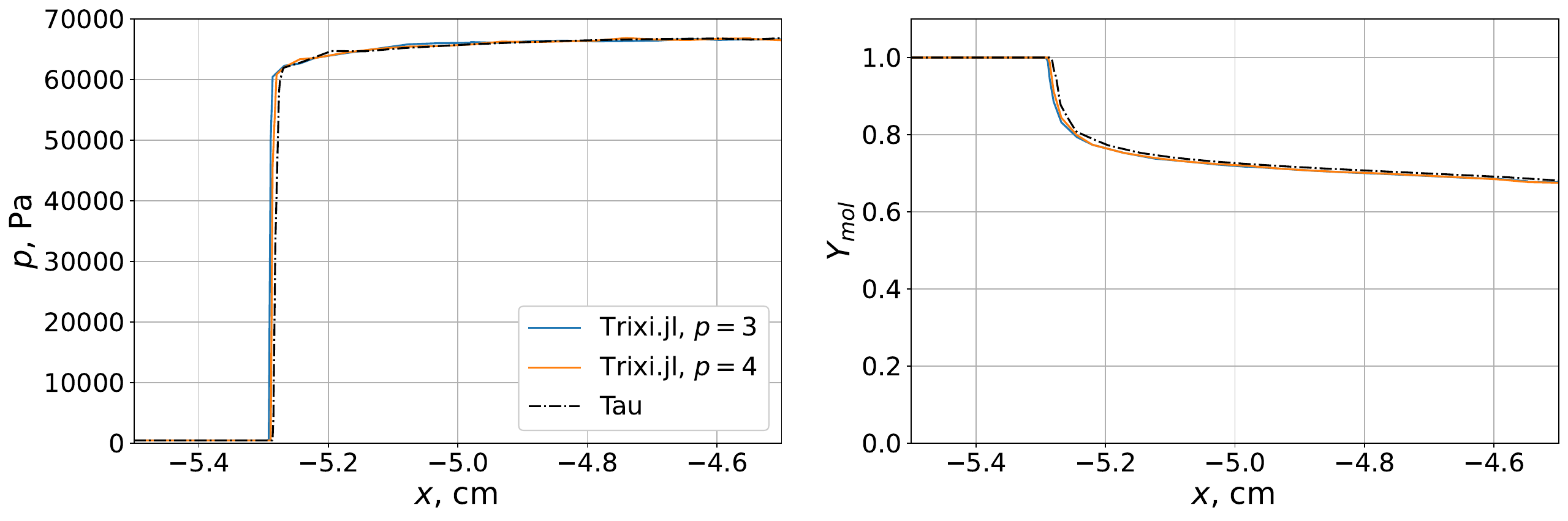}
  \caption{Pressure (left) and mass fraction of molecular oxygen (right) along the stagnation line computed with the DG method with the entropy-conservative flux (solid lines) and with the DLR TAU solver (dash-dot line).}
  \label{fig:TrixiTausline}
\end{figure}

Figure~\ref{fig:TrixiTau} shows the fields of the pressure $p$ and the mass fraction of molecular oxygen $Y_{\mathrm{O}_2}$ computed using the DG approach presented in the present paper and using the DLR TAU solver as a reference solution. Excellent qualitative agreement between the solvers can be seen throughout the whole domain. For a more detailed comparison, we plot the pressure and the component mass fractions along the stagnation line, as shown on Figure~\ref{fig:TrixiTausline}. Simulation results obtained using 3-rd and 4-th order polynomials in the DG method are shown, and excellent agreement can be seen between the DG results and those obtained with the DLR TAU solver, with only some very minor differences near the shock.

\section{Conclusions}
In the present work, we have developed a numerical algorithm for the computation of entropy-conservative fluxes for gases
with internal degrees of freedom for use with the DGSEM method. We verified the developed approach by considering a non-reacting test case of a weak blast wave and the spatially homogeneous
chemical relaxation of an oxygen mixture. Finally, we applied the developed framework to simulation of a Mach 10 flow around a cylinder, and
compare the results with those obtained with the DLR TAU code.
Excellent agreement is observed between the results, thus verifying the developed approach and showing the possibility of using
high-order entropy-stable methods for high-enthalpy flows with strong shocks and non-conforming meshes obtained via AMR.

\section{Acknowledgments}
This work has been supported by the German Research Foundation within the research unit SFB 1481 and the research unit DFG-FOR5409.

\end{document}